\documentclass{llncs}
\usepackage{amsfonts}
\usepackage{amsmath}
\def\smallspace{\:\!}
\newcommand{\innerop}[3]{\mbox{$\langle #1 | #2 | #3 \rangle$}}
\newcommand{\hps}[2]{\makebox[0pt][l]{$#1$}\phantom{#2}}
\newcommand{\hpsr}[2]{\phantom{#2}\makebox[0pt][r]{$#1$}}

\title{Quantum Counting}

\author{%
Gilles Brassard\,\inst{1}\thanks{\,Supported in
part by Canada's {\sc nserc}, Qu\'ebec's {\sc fcar} and the Canada Council.}
\and
Peter H{\o}yer\,\inst{2}\thanks{\,Supported in part by the {\sc esprit}
Long Term Research Programme of the EU under project number 20244
({\sc alcom-it}). Research carried out while this author was at the
Universit\'e de Montr\'eal.}
\and
Alain Tapp\,\inst{1}\thanks{\,Supported in
part by  postgraduate fellowships from {\sc fcar}
and  {\sc nserc}.}}

\institute{Universit\'e de Montr\'eal,
\email{\textnormal{\{}brassard\textnormal{,}%
tappa\textnormal{\}}@iro.umontreal.ca}
\and Odense University, \email{u2pi@imada.ou.dk}}

\begin{document}

\def\squareforqed{\hbox{\rlap{$\sqcap$}$\sqcup$}}
\def\qed{\ifmmode\squareforqed\else{\unskip\nobreak\hfil
\penalty50\hskip1em\null\nobreak\hfil\squareforqed
\parfillskip=0pt\finalhyphendemerits=0\endgraf}\fi}

\newcommand{\fl}[1]{\mbox{$\lfloor \mbox{$#1$} \rfloor$}}
\newcommand{\ket}[1]{\mbox{$| #1 \rangle$}}
\newcommand{\braket}[1]{\mbox{$\langle #1 | #1 \rangle$}}
\def\integer{\mathbb{Z}}
\newcommand{\pbfrac}[2]{\mbox{$\mbox{}^{#1}\!/_{#2}$}}
\newcommand{\pbdemi}{\mbox{$\textstyle \pbfrac{1}{2}$}}
\newcommand{\ph}[2]{\vphantom{#1}\smash{#2}}

\newcommand{\novert}{\smash{N/t}\vphantom{N}}
\newcommand{\tovern}{\smash{t/N}\vphantom{N}}
\newcommand{\PT}{\mbox{\bf P}}
\newcommand{\NP}{\mbox{\bf NP}}
\newcommand{\SP}{\mbox{\bf \#P}}
\newcommand{\npc}{\mbox{\NP--com}\-plete}
\newcommand{\spc}{\mbox{\SP--com}\-plete}
\newcommand{\BQP}{\mbox{\bf BQP}}
\newcommand{\bigket}[1]{\mbox{$\left| #1 \right\rangle$}}
\newcommand{\jqa}{\,.\,.\,}
\newcommand{\half}{\frac{1}{2}}

\newcommand{\Nr}{\mbox{\vphantom{$N$}\smash{$N/r$}}}
\newcommand{\Nt}{\mbox{\vphantom{$N$}\smash{$N/t$}}}

\newcommand{\mb}[1]{\mbox{$#1$}}

\maketitle

\begin{abstract}
We study some extensions of
Grover's quantum searching \mbox{algorithm}.
First, we generalize the Grover iteration in the
light of a concept called amplitude amplification.
Then, we show that the quadratic speedup obtained by the quantum searching
algorithm over classical brute force can still be obtained for
a large family of search problems for which good classical
heuristics exist.
Finally, as our main result,
we combine ideas from Grover's and Shor's quantum algorithms
to perform approximate counting, which can be seen as
an amplitude estimation process.
\looseness=-1
\end{abstract}

\section{Introduction}
Quantum computing is a field at the junction
of theoretical modern physics and
theoretical computer science.
Practical experiments involving a few quantum bits
have been successfully performed, and
much progress has been achieved
in quantum information theory, quantum error correction
and fault tolerant quantum computation.
Although we are still far from
having desktop quantum computers in our offices,
the quantum computational paradigm
could soon be more than mere theoretical
exercise~\cite[and references therein]{horizon}.

The discovery by Peter Shor~\cite{Shor97} of a polynomial-time
quantum algorithm for factoring and computing discrete logarithms
was a major milestone in the history of quantum computing.
Another significant result is Lov Grover's quantum search
algorithm~\cite{Grover97}.
Grover's algorithm does not solve \npc{}
problems in polynomial time,
but the wide range of its applications
compensates for this.

The search problem and Grover's iteration are reviewed in
Section~\ref{sec:amplifi}. It was already implicit in \cite{BH97}
that the heart of Grover's algorithm can be viewed as an amplitude
amplification process. Here, we develop this
viewpoint and obtain a more general algorithm.

When the structure in a search problem cannot be exploited,
any quantum algorithm
requires a computation time at least proportional to the square root
of the time taken by brute-force classical searching~\cite{BBBV97}.
In~practice, the structure of the search problem
can usually  be exploited,
yielding deterministic or heuristic
algorithms that are much more efficient than brute force would be.
In~Section~\ref{sec:heuri}, we study a vast family of heuristics
for which we show how to adapt the quantum search algorithm
to preserve quadratic speedup over classical techniques.

In~Section~\ref{sec:count},
we present, as our main result, a quantum algorithm to perform counting.
This is the problem of counting the number of elements
that fulfill some specific requirements,
instead of merely finding such an element.
Our algorithm builds on both Grover's iteration~\cite{Grover97}
as described
in~\cite{BBHT96} and the quantum Fourier transform as used
in~\cite{Shor97}.
The accuracy of the algorithm
depends on the amount of time one is willing to invest.
As Grover's algorithm is a special case of the  amplitude amplification
process, our counting algorithm can also be viewed as a special case
of the more general process of {\em amplitude estimation}.

We~assume in this paper that the reader is familiar with
basic notions of quantum
computing \mbox{\cite{Barenco96,Brassard96}}.

\section{Quantum Amplitude Amplification} \label{sec:amplifi}
Consider the following search problem:
Given a Boolean function $F : X \rightarrow \{0,1\}$ defined on some
finite domain~$X$, find an input $x \in X$ for which $F(x)=1$, provided such
an $x$ exists.
We assume that $F$ is given as a black~box, so that it is not
possible to obtain knowledge about $F$ by any other means than evaluating
it on points in its domain.
The best classical strategy is to evaluate $F$ on
random elements of~$X$.
If~there is a unique $x_0 \in X$ on which $F$ takes value~1,
this strategy evaluates $F$ on
roughly half the elements of the domain in order to determine~$x_0$.
By~contrast, Grover~\cite{Grover97} discovered
a quantum algorithm that only requires an expected number
of evaluations of $F$ in the order of~$\sqrt{N}\,$,
where $N=|X|$ denotes the cardinality of~$X$.

It is useful for what follows to think of the above-mentioned classical
strategy in terms of an algorithm that keeps boosting the
probability of finding~$x_0$.
The algorithm evaluates $F$ on new inputs, until it
eventually finds the unique input~$x_0$ on which $F$ takes value~1.
The probability that the algorithm stops after exactly $j$ evaluations
of $F$ is~$1/N$ ($1 \leq j \leq N-2$), and thus
we can consider that each evaluation boosts the
probability of success by an additive amount of~$1/N$.

Intuitively, the quantum analog of boosting the probability of success
would be to boost the {\em amplitude\/} of being in a certain subspace
of a Hilbert space, and indeed the algorithm found by Grover can be
seen as working by that latter principle~\cite{Grover97,BBHT96}.
As~discovered by Brassard and H{\o}yer~\cite{BH97}, the idea of
amplifying the amplitude of a subspace is a technique that applies
in general.  Following~\cite{BH97}, we refer to this as
{\em amplitude amplification}, and describe the technique below.
For this, we require the following notion, which we shall use throughout
the rest of this section.

Let $\ket{\Upsilon}$ be any pure state of a joint quantum
system~$\mathcal H$.
Write $\ket{\Upsilon}$ as a superposition of orthonormal
states according to the state of the first subsystem:
\[\ket{\Upsilon} = \sum_{i \in \integer} x_i \ket{i}\ket{\Upsilon_i} \]
so that only a finite number of the states \ket{i}\ket{\Upsilon_i}
have nonzero amplitude~$x_i$.

Every Boolean function $\chi : \integer \rightarrow \{0,1\}$
induces two orthogonal subspaces of~$\mathcal H$, allowing
us to rewrite \ket{\Upsilon} as follows:
\begin{equation}\label{eq:partioning}
\ket{\Upsilon} = \ket{\Upsilon^a} + \ket{\Upsilon^b}
    = \sum_{i \in \chi^{-1}(1)} x_i \ket{i}\ket{\Upsilon_i}
    + \sum_{i \in \chi^{-1}(0)} x_i \ket{i}\ket{\Upsilon_i}.
\end{equation}
We say that a state \ket{i}\ket{\cdot} is {\em good\/} if $\chi(i)=1$,
and otherwise it is {\em bad}.
Thus, we have that \ket{\Upsilon^a} denotes the
projection of \ket{\Upsilon} onto the subspace spanned by the good states,
and similarly \ket{\Upsilon^b} is the projection of \ket{\Upsilon}
onto the subspace spanned by the bad states.
Let $a_\Upsilon = \braket{\Upsilon^a}$ denote the probability that
measuring \ket{\Upsilon} produces a good state, and similarly
let $b_\Upsilon = \braket{\Upsilon^b}$.
Since \ket{\Upsilon^a} and \ket{\Upsilon^b} are orthogonal,
we have $a_\Upsilon+b_\Upsilon=1$.

Let $\mathcal A$ be any quantum algorithm that acts on~$\mathcal H$ and
uses no measurements.  The heart of amplitude amplification is the following
operator~\cite{BH97}
\begin{equation}\label{eq:defq}
{\mathbf Q} = {\mathbf Q}({\mathcal A},\chi,\phi,\varphi)
  = - {\mathcal A} \smallspace {\mathbf S}_0^\phi \smallspace
  {\mathcal A}^{-1} \smallspace {\mathbf S}_\chi^\varphi.
\end{equation}
Here, $\phi$ and~$\varphi$ are complex numbers of unit norm, and
operator ${\mathbf S}_\chi^\varphi$ conditionally changes the phase
by a factor of $\varphi$:
\[\ket{i}\ket{\cdot} \,\longmapsto\, \begin{cases}
  \varphi \ket{i}\ket{\cdot} & \text{ if $\chi(i) = 1$}\\
  \hphantom{\varphi}\ket{i}\ket{\cdot} & \text{ if $\chi(i) = 0$.}
\end{cases}\]
Further, ${\mathbf S}_0^\phi$ changes the phase of a state
by a factor of~$\phi$ if and only if the first register holds a zero.
The operator~$\mathbf Q$ is a generalization of the iteration
applied by Grover in his original quantum searching paper~\cite{Grover97}.
It was first used in~\cite{BH97} to obtain an exact quantum
polynomial-time algorithm for Simon's problem.
It~is well-defined since we assume that $\mathcal A$ uses
no measurements and, therefore, $\mathcal A$~has an inverse.

Denote the complex conjugate of $\lambda$ by~$\lambda^*$.
It is easy to show the following lemma by a few simple rewritings.
\begin{lemma}\label{lm:backforth}
Let $\ket{\Upsilon}$ be any superposition.  Then
\[{\mathcal A} \smallspace {\mathbf S}_0^\phi
  \smallspace {\mathcal A}^{-1}
  \ket{\Upsilon} \,=\, \ket{\Upsilon}
  - (1-\phi) \smallspace {\innerop{\Upsilon}{\mathcal A}{\mathbf 0}}^*
    \smallspace {\mathcal A} \,\ket{\mathbf 0}.\]
\end{lemma}

By~factorizing $\mathbf Q$ as
$({\mathcal A} \smallspace {\mathbf S}_0^\phi \smallspace
{\mathcal A}^{-1}) (-{\mathbf S}_\chi^\varphi)$, the next lemma follows.

\begin{lemma}\label{lm:fourdim}
Let $\ket{\Upsilon} = \ket{\Upsilon^a} + \ket{\Upsilon^b}$
be any superposition.  Then
\begin{align}
{\mathbf Q} \,\ket{\Upsilon^a} &= -\varphi \ket{\Upsilon^a}
  + \varphi (1-\phi) \smallspace
 {\innerop{\Upsilon^a}{\mathcal A}{\mathbf 0}}^*
 \smallspace {\mathcal A} \ket{\mathbf 0}\label{eq:fourdim1}\\
 {\mathbf Q} \,\ket{\hps{\Upsilon^b}{\Upsilon^a}} &=
  \hpsr{-}{-\varphi} \ket{\hps{\Upsilon^b}{\Upsilon^a}}
  + \hphantom{\varphi} (1-\phi) \smallspace
 {\innerop{\hps{\Upsilon^b}{\Upsilon^a}}{\mathcal A}{\mathbf 0}}^*
 \smallspace {\mathcal A} \ket{\mathbf 0}.\label{eq:fourdim2}
\end{align}
\end{lemma}

In particular, letting $\ket{\Upsilon}$ be
${\mathcal A} \ket{\mathbf 0} = \ket{\Psi^a} + \ket{\Psi^b}$
implies that the subspace spanned by
\ket{\Psi^a} and \ket{\Psi^b} is invariant
under the action of~$\mathbf Q$.

\begin{lemma}\label{lm:twodim}
Let ${\mathcal A} \ket{\mathbf 0} = \ket{\Psi}
= \ket{\Psi^a} + \ket{\Psi^b}$.  Then
\begin{align}
{\mathbf Q} \,\ket{\Psi^a}
&= \hpsr{\varphi}{-} ((1-\phi)a-\hps{1}{\phi}) \ket{\Psi^a}
+ \hpsr{\varphi(1-\phi)a}{(1-\phi)(1-a)}
  \ket{\hps{\Psi^b}{\Psi^a}}\label{eq:twodim1}\\
{\mathbf Q} \,\ket{\hps{\Psi^b}{\Psi^a}}
&= -((1-\phi)a+\phi) \ket{\hps{\Psi^b}{\Psi^a}}
+ (1-\phi)(1-a) \ket{\Psi^a},\label{eq:twodim2}
\end{align}
where $a=\braket{\Psi^a}$.
\end{lemma}

\mbox{From} Lemmas~\ref{lm:fourdim} and~\ref{lm:twodim} it follows that,
for any vector $\ket{\Upsilon} = \ket{\Upsilon^a} + \ket{\Upsilon^b}$,
the subspace spanned by the set
$\{\ket{\Upsilon^a}, \ket{\Upsilon^b}, \ket{\Psi^a}, \ket{\Psi^b}\}$
is invariant under the action of~$\mathbf Q$.
By~setting $\phi = \varphi = -1$, we find the following
much simpler expressions.

\begin{lemma}\label{lm:simpler}
Let ${\mathcal A} \ket{\mathbf 0} = \ket{\Psi}
= \ket{\Psi^a} + \ket{\Psi^b}$, and let
${\mathbf Q} = {\mathbf Q}({\mathcal A},\chi,-1,-1)$.
Then
\begin{align}
{\mathbf Q} \,\ket{\Psi^a}
&= (1-2a) \ket{\Psi^a} - 2a \ket{\hps{\Psi^b}{\Psi^a}}
\label{eq:simpler1}\\
{\mathbf Q} \,\ket{\hps{\Psi^b}{\Psi^a}}
&= (1-2a) \ket{\hps{\Psi^b}{\Psi^a}} + 2\hps{b}{a} \ket{\Psi^a},
\label{eq:simpler2}
\end{align}
where $a=\braket{\Psi^a}$ and $b=1-a=\braket{\Psi^b}$.
\end{lemma}

The recursive formulae defined by Equations~\ref{eq:simpler1}
and~\ref{eq:simpler2} were solved in~\cite{BBHT96}, and
their solution is given in the following theorem.
The general cases defined by
Equations~\mbox{\ref{eq:fourdim1}\;\!--\;\!\ref{eq:twodim2}}
have similar solutions, but we shall not
need them in what follows.

\begin{theorem}[Amplitude Amplification---simple case]\label{thm:simpler}
Let ${\mathcal A} \ket{\mathbf 0} = \ket{\Psi}
= \ket{\Psi^a} + \ket{\Psi^b}$, and let
${\mathbf Q} = {\mathbf Q}({\mathcal A},\chi,-1,-1)$.
Then, for all $j \geq 0$,
\[{\mathbf Q}^j {\mathcal A} \, \ket{\mathbf 0}
  = k_j \ket{\Psi^a} + \ell_j \ket{\Psi^b},\]
where
\[k_j = \frac{1}{\sqrt{a\;\!}}\sin((2j+1)\theta) \quad \text{ and }
  \quad \ell_j = \frac{1}{\sqrt{1-a\;\!}}\cos((2j+1)\theta),\]
and where $\theta$ is defined so that
$\sin^2 \theta = a = \braket{\Psi^a}$ and $0 \leq \theta \leq \pi/2$.
\end{theorem}

Theorem~\ref{thm:simpler} yields a method for boosting the
success probability~$a$ of a quantum algorithm~$\mathcal A$.
Consider what happens if we apply $\mathcal A$ on the initial
state \ket{\mathbf 0} and then measure the system.
The probability that the outcome is a good state is~$a$.
If,~instead of applying~$\mathcal A$,
we apply operator ${\mathbf Q}^m {\mathcal A}$
for some integer $m \geq 1$,
then our success probability is given by $a k_m^2 = \sin^2((2m+1)\theta)$.
Therefore, to obtain a high probability of success, we want to
choose integer~$m$ such that $\sin^2((2m+1)\theta)$ is close to~1.
Unfortunately, our ability to choose $m$ wisely depends on our
knowledge about~$\theta$, which itself depends on~$a$.
The two extreme cases are when we know the exact
value of~$a$, and when we have no prior knowledge about $a$ whatsoever.

Suppose the value of~$a$ is known.
If $a > 0$, then by letting $m = \lfloor \pi /4 \theta \rfloor$,
we have that $a k_m^2 \geq 1 -a$, as shown in~\cite{BBHT96}.
The next theorem is immediate.

\begin{theorem}[Quadratic speedup]\label{thm:quad}
Let $\mathcal A$ be any quantum algorithm that uses no measurements,
and let $\chi : \integer \rightarrow \{0,1\}$ be any Boolean function.
Let the initial success probability~$a$ and angle~$\theta$ be
defined as in Theorem~\ref{thm:simpler}.
Suppose $a>0$ and set $m = \lfloor \pi /4 \theta \rfloor$.
Then, if we compute ${\mathbf Q}^m {\mathcal A} \ket{\mathbf 0}$
and measure the system, the outcome is good with
probability at least~\mbox{$\max(1-a,a)$}.
\end{theorem}

This theorem is often referred to as a quadratic speedup,
or the square-root running-time result.  The reason for this is that if
an algorithm ${\mathcal A}$ has success probability~$a>0$,
then after an expected number of $1/a$ applications of~$\mathcal A$,
we will find a good solution.
Applying the above theorem reduces this to an expected number of
at most $(2m+1)/(1-a) \in \Theta(\sqrt{1/a\;\!}\;\!)$ applications
of $\mathcal A$ and its inverse.

Suppose the value of $a$ is known and that \mbox{$0<a<1$}.
Theorem~\ref{thm:quad} allows us to find a good solution with
probability at least~$\max(1-a,a)$.
A~natural question to ask is whether it is possible to improve this
to certainty, still given the value of~$a$.
It turns out that the answer is positive.
This is unlike classical computers,
where no such general de-randomization technique is known.
We now describe two optimal methods for obtaining this,
but other approaches are possible.

The first method is by applying amplitude amplification, not on the
original algorithm~$\mathcal A$, but on a slightly modified version
of it.  If ${\tilde m} = \pi / 4 \theta - 1/2$ is an integer, then
we would have $\ell_{\tilde m} = 0$, and we would succeed with
certainty.
In~general, $m_0 = \lceil {\tilde m}\rceil$ iterations is
a fraction of 1~iteration too many,
but we can compensate for that by choosing
$\theta_0 = \pi/(4 m_0 +2)$, an angle slightly smaller than~$\theta$.
Any quantum algorithm that succeeds
with probability~$a_0$ such that \mbox{$\sin^2 \theta_0 = a_0$},
will succeed with certainty after $m_0$ iterations of
amplitude amplification.
Given $\mathcal A$ and its initial success probability~$a$,
it is easy to construct a new quantum algorithm that succeeds
with probability \mbox{$a_0 \leq a$}:
Let $\mathcal B$ denote the quantum algorithm that takes a single
qubit in the initial state \ket{0} and
rotates it to the superposition
$\sqrt{1-a_0/a\;\!}\;\! \ket{0} + \sqrt{a_0/a\;\!}\;\! \ket{1}$.
Apply both $\mathcal A$ and $\mathcal B$, and define
a good solution as one in which $\mathcal A$ produces a good
solution, and the outcome of $\mathcal B$ is the state~\ket{1}.

The second method
is to slow down the speed of the very
last iteration.  First, apply $m_0 = \lfloor{\tilde m}\rfloor$
iterations of amplitude amplification with $\phi = \varphi = -1$.
Then, if~$m_0 < \tilde m$,
apply one more iteration with complex phase-shifts $\phi$ and $\varphi$
satisfying $\ell_{m_0}^2 = 2 a (1 -\textup{Re}(\phi))$ and so
that $\varphi(1-\phi)a k_{m_0} - ((1-\phi)a + \phi) \ell_{m_0}$
vanishes.  Going through the algebra and applying
Lemma~\ref{lm:twodim} shows that this
produces a good solution with certainty.
For the case $m_0 = 0$, this second method was independently
discovered by Chi and Kim~\cite{CK98}.

Suppose now that the value of~$a$ is not known.
In~Section~\ref{sec:count}, we discuss techniques for finding a
good estimate of~$a$, after which one then can apply a
weakened version of Theorem~\ref{thm:quad} to find a good solution.
Another idea is to try to find a good solution without prior computation
of an estimate of~$a$.  Within that approach,
by adapting the ideas in Section~4 in~\cite{BBHT96}
(Section~6 in its final version),
we can still obtain a quadratic speedup.

\begin{theorem}[Quadratic speedup without knowing~{\boldmath $a$}]
\label{thm:withouta}
Let $\mathcal A$ be any quantum algorithm that uses no measurements,
and let $\chi : \integer \rightarrow \{0,1\}$ be any Boolean function.
Let the initial success probability~$a$ of $\mathcal A$
be defined as in Theorem~\ref{thm:simpler}.
Then there exists a quantum algorithm that
finds a good solution using an expected number
of $\Theta(\sqrt{1/a\;\!}\;\!)$ applications of $\mathcal A$ and
its inverse if $a>0$, and otherwise runs forever.
\end{theorem}

By applying this theorem to the searching problem defined
in the first paragraph of this section,
we obtain the following result from~\cite{BBHT96},
which itself is a generalization of the work by Grover~\cite{Grover97}.

\begin{corollary}\label{cor:search}
Let $F:X \rightarrow \{0,1\}$ be any Boolean function defined on
a finite set~$X$.
Then there exists a quantum algorithm {\bf Search} that finds
an $x \in X$ such that $F(x)=1$ using an expected number
of $\Theta(\sqrt{|X|/t\;\!}\;\!)$ evaluations of~$F$,
provided such an $x$ exists, and otherwise runs forever.
Here \mbox{$t = |\{x \in X \mid F(x)=1\}|$}
denotes the cardinality of the preimage of~1.
\end{corollary}

\begin{proof}
Apply Theorem~\ref{thm:withouta} with \mbox{$\chi = F$} and
$\mathcal A$ being any unitary transformation that maps
\ket{0} to $\frac{1}{\sqrt{|X|\;\!}\;\!} \sum_{x \in X} \ket{x}$,
such as the Walsh--Hadamard transform.
\qed
\end{proof}

\section{Quantum Heuristics}\label{sec:heuri}

If function $F$
has no useful structure, then quantum algorithm {\bf Search}
will be more efficient than any classical
(deterministic or probabilistic) algorithm.
In~sharp contrast, if some useful information is known
about the function, then some classical algorithm might be
very efficient.
Useful information might be clear mathematical statements
or intuitive information stated as a probability distribution
of the likelihood of $x$ being a solution.
The information we have about $F$ might also be expressed as
an efficient classical heuristic to find a solution.
In~this section, we address the problem of heuristics.

Search problems, and in particular \NP{} problems, are often
very difficult to solve.  For many \npc{} problems,
practical algorithms are known that are more
efficient than brute force search on the average: they
take advantage of the problem's structure
and especially of the input distribution.
Although in general very few theoretical results
exist about the efficiency of heuristics,
they are very efficient in practice.

We concentrate on a large but simple family of heuristics
that can be applied to search problems.
Here, by heuristics, we mean a probabilistic algorithm running
in polynomial time that outputs what one is searching for with
some nonzero probability.
Our goal is to apply Grover's technique
for heuristics in \mbox{order} to speed them up, in the same way that
Grover speeds up black-box search, without
making things too complicated.

More formally, suppose we have a family~${\mathcal F}$ of functions
such that each \mbox{$F\in {\cal F}$} is of the form
$F:X \rightarrow \{0,1\}$.
A~{\em heuristic\/} is a function
$G: {\mathcal F} \times R \rightarrow X$, for an appropriate
finite set~$R$.
For every function $F \in {\mathcal F}$, let $t_F=|F^{-1}(1)|$ and
\mbox{$h_F=|\{r \in R \mid F(G(F,r))=1\}|$}.
We say that the heuristic is {\em efficient\/} for a given $F$ if
$h_F/|R|  > t_F/|X|$ and the heuristic is {\em good\/} in general if
\[\textup{E}_{\cal{F}}\left( \frac{h_F}{|R|} \right) \ > \
\textup{E}_{\cal{F}}\left( \frac{t_F}{|X|} \right) \ .\]
Here $\textup{E}_{\cal{F}}$ denotes the expectation over all $F$ according
to some fixed distribution.
Note that for some $F$, $h_F$ might be small but
repeated uses of the heuristic,
with seeds $r$ uniformly chosen in $R$,
will increase the probability of finding a solution.

\begin{theorem}
Let $F$ be a search problem chosen in a family ${\cal F}$
according to some probability distribution.  If,~using a heuristic $G$,
a solution to $F$ is found in \mbox{expected} time $T$
then, using a quantum computer,
a solution can be found in \mbox{expected} time
in~$O(\sqrt{T}\,)$.
\end{theorem}

\begin{proof}
We simply combine the quantum algorithm {\bf Search} with the heuristic~$G$.
Let $G'(r) = F(G(F,r))$ and
$x = G(F,\textup{\textbf{Search}}(G'))$,
so that $F(x)=1$.
By~Corollary~\ref{cor:search},
for each function
$F \in {\cal F}$, we have an
expected running time in $\Theta(\sqrt{|R|/h_F}\,)$.
Let $P_F$ denote the probability that $F$ occurs.
Then \mbox{$\sum_{F \in \cal{F}} P_F =1$}, and
we have that the expected running time is
in the order of
$\sum_{F\in {\cal F}} \sqrt{|R|/h_F\;\!}\;\! P_F$, which can be
rewritten as
\[\sum_{F\in {\cal F}}   \sqrt{\frac{|R|}{h_F} P_F} \sqrt{P_F}
 \leq    \left(\sum_{F\in {\cal F}} \frac{|R|}{h_F} P_F \right)^{1/2}
\left(\sum_{F\in {\cal F}} P_F \right)^{1/2}
  =   \left(\sum_{F\in {\cal F}} \frac{|R|}{h_F} P_F \right)^{1/2}, \]
by Cauchy--Schwarz's inequality.
\qed
\end{proof}

\section{Approximate Counting}\label{sec:count}

In this section, we do not concentrate on finding one solution,
but rather on counting them.
For~this, we complement Grover's iteration~\cite{Grover97} using
techniques inspired by Shor's quantum factoring algorithm~\cite{Shor97}.

\medskip\noindent
{\bf Counting Problem}:
Given a Boolean function $F$
defined on some finite set $X=\{0,\dots,N-1\}$,
find or approximate
$t=\big| F^{-1}(1) \big|$.
\medskip

Before we proceed, here is the basic intuition.
\mbox{From} Section~\ref{sec:amplifi}
it follows that, in Grover's algorithm,
the amplitude of the set $F^{-1}(1)$,
as well as the amplitude of the set $F^{-1}(0)$,
varies with the number of
iterations according to a periodic function.
We also note that the period (frequency) of this association is
in {\em direct} relation with the sizes of these sets.
Thus, estimating their common period using Fourier analysis will
give us useful information on
the sizes of those two sets.
Since the period will be the same if $F^{-1}(1)$
has cardinality $t$ or if $F^{-1}(1)$ has cardinality $N-t$,
we will assume in the rest of this section that $t \leq N/2$.

The quantum algorithm {\bf Count} we give to solve this problem
has two parameters: the function $F$ given
as a black box and an integer $P$ that will determine the precision
of our estimate, as well as the time taken by the algorithm.
For simplicity,
we assume that $P$ and $N$ are powers of 2,
but this is not essential.
Our algorithm~is based on the following two unitary transformations:
\begin{gather*}
{\mathbf C}_F :\; \ket{m} \otimes \ket{\Psi} \;\;\rightarrow\;\;
              \ket{m} \otimes ({\mathbf G}_F)^m \ket{\Psi}\\
{\mathbf F}_P \;:\; \ket{k} \;\;\rightarrow\;\; \frac{1}{\sqrt{P\;\!}}
              \sum^{P-1}_{l=0} e^{2 \pi \imath kl/P} \,\ket{l}.
\end{gather*}
Here $\imath = \sqrt{-1}\;\!$ and
${\mathbf G}_F = {\mathbf Q}({\mathbf W},F,-1,-1)$ denotes
the iteration originally used by Grover~\cite{Grover97},
where ${\mathbf W}$ denotes the Walsh--Hadamard transform
on $n$~qubits that maps
\ket{0} to $2^{-n/2} \sum_{i=0}^{2^n-1}\ket{i}$.

In order to apply ${\mathbf C}_F$ even if its first argument is in a quantum
superposition, it is necessary to have an upper bound on the value of~$m$,
which is the purpose of parameter~$P$.
Thus, unitary transformation ${\mathbf C}_F$ performs exactly $P$ Grover's
iterations so that $P$ evaluations of $F$ are required.
The quantum Fourier transform can be efficiently implemented
(see~\cite{Shor97} for example).

\bigskip\noindent $\textbf{\bf Count}(F,P)$
\begin{enumerate}
\item $\ket{\Psi_0}  \ \leftarrow \  {\mathbf W} \otimes {\mathbf W}
       \;\ket{0}\ket{0} $
\item  $\ket{\Psi_1}  \ \leftarrow \ {\mathbf C}_F \,\ket{\Psi_0} $
\item  $\ket{\Psi_2}  \ \leftarrow \  \ket{\Psi_1}$ after the second
register is measured ({\em optional}\,) \label{stp:measure}
\item  $\ket{\Psi_3}  \ \leftarrow \ {\mathbf F}_P \otimes {\mathbf I}
       \;\ket{\Psi_2} $ \label{stp:fourier}
\item  $\tilde{f}  \ \leftarrow \  $  measure $\ket{\Psi_3}$
       \quad\quad\quad (if $\tilde{f} >P/2$
       then $\tilde{f} \leftarrow (P-\tilde{f})$)
\item  output: $\ N \sin^2(\tilde{f} \pi/P)$
       \quad\ \;(and $\tilde f$ if needed)
\end{enumerate}

The following theorem tells us how to make proper use of algorithm
{\bf Count}.

\begin{theorem}\label{thm:count}
Let $F:\{0,\ldots,N-1\}\rightarrow\{0,1\}$ be a Boolean function,
\mbox{$t=|F^{-1}(1)| \leq N/2$} and $\tilde{t}$ be the output of
$\textup{\textbf{Count}}(F,P)$ with $P \geq 4$, then
\[|t - \tilde{t}| < \frac{2 \pi}{P} \sqrt{tN} + \frac{\pi^2}{P^2} N\]
with probability at least  $8/\pi^2$.
\end{theorem}

\begin{proof}
Let us follow the state through the algorithm using notation from
\mbox{Section}~\ref{sec:amplifi}.
\begin{align*}
\ket{\Psi_0} &= \frac{1}{\sqrt{PN\;\!}\;\!} \sum_{m=0}^{P-1}
   \sum_{x =0}^{N-1}
   \ket{m}\ket{x} \\
\ket{\Psi_1} &= \frac{1}{\sqrt{P\;\!}\;\!} \sum_{m=0}^{P-1}
    \ket{m} \Bigg( k_m \sum_{x \in F^{-1}(1)} \ket{x}
   \ + \ \ell_m \sum_{x \in F^{-1}(0)} \ket{x}  \Bigg).
\end{align*}
We introduced Step~\ref{stp:measure} to make it intuitively clear
to the reader why the Fourier transform in Step~4 gives us what we
want. The result of this measurement is not used in the algorithm and this is
why it is optional:
the final outcome would be the same if Step~\ref{stp:measure}
were not performed.
Without loss of generality, assume that the state $x$ observed in the
second register is such that $F(x)=1$.
Then by replacing $k_m$ by its definition we obtain
\begin{equation}
\ket{\Psi_2} = \alpha \sum_{m=0}^{P-1}\sin((2m+1) \theta)
  \,\ket{m},
\end{equation}
where $\alpha$ is a normalization factor that depends on~$\theta$.

Let
\begin{equation}\label{eqn:f}
f= P \theta / \pi.
\end{equation}
In Step~\ref{stp:fourier},
we apply the Fourier transform on a sine (cosine) of period $f$ and
phase shift $\theta$.
\mbox{From} $\sin^2 \theta =t/N$  we conclude that $\theta \leq \pi/2$
and $f \leq P/2$.
After we apply the Fourier transform, the state $\ket{\Psi_3}$
strongly depends on $f$ (which depends on $t$).
If $f$ were an integer, there would be two possibilities:
either $f=0$ (which happens if $t=0$ or $t=N$),
in which case
$\ket{\Psi_3} =  \ket{0}$,
or $t>0$, in which case
$\ket{\Psi_3} =  a\ket{f} + b\ket{P-f}$,
where $a$ and $b$ are complex numbers of norm $1/\sqrt{2}$.

In general $f$ is not an integer and we will obtain something more
complicated.  We define $f^-=\fl{f}$ and $f^+ = \fl{f+1}$.
We still have three cases.
If $1<f<P/2-1$, we obtain
\[ \ket{\Psi_3} =    a\ket{f^-} + b\ket{f^+}  + c\ket{P-f^-}
    + d\ket{P-f^+} + \ket{R} \]
where $\ket{R}$ is an un-normalized error term that may include
some or all values other than the desirable $f^-$, $f^+$, \mbox{$P-f^-$}
and \mbox{$P-f^+$}.
The two other possibilities are $0<f<1$, in which case we obtain
\[ \ket{\Psi_3}  =  a\ket{0} + b\ket{1}  + c\ket{P-1} +  \ket{R} \]
or $P/2-1<f<P/2$, in which case we obtain
\[ \ket{\Psi_3} = a\ket{P/2-1} + b\ket{P/2}  + c\ket{P/2+1} + \ket{R} \, .\]
In all three cases, extensive algebraic manipulation shows that
the square of the norm of the error term $\ket{R}$ can be
upper bounded by~$2/5$,
\[ \braket{R} < \frac{2}{5} \, . \]
In order to bound the success probability by $8/\pi^2$
(which is roughly $0.81$ and therefore larger than $1-2/5=0.6$)
as claimed in the statement of the Theorem,
we could perform a complicated case analysis depending
on whether the value $x$ observed in Step~\ref{stp:measure} is such that
$F(x)=0$ or $F(x)=1$.
Fortunately, in the light of some recent
analysis of Michele Mosca \cite{Mo98}, which itself is based on
results presented in \cite{CEMM98},
this analysis can be simplified.
Since the information obtained by measuring the second
register is not used, measuring it in a different basis would
not change the behaviour of the algorithm.
Measuring in the eigenvector basis of ${\mathbf G}_F$,
one obtains this bound in an elegant way.
Details will be provided in the final version of this paper.

Assuming that $\tilde{f}$ has been observed at Step~5
and applying Equation~\ref{eqn:f} and the fact that
$\sin \theta = \sqrt{t/N}$,
we obtain an estimate
$\tilde{t}$ of $t$ such that
\[|t - \tilde{t}| < \frac{2 \pi}{P} \sqrt{tN} + \frac{\pi^2}{P^2} N \, . \]
\qed
\end{proof}

Using a similar technique, it can be shown that
the same quantum algorithm can also be used to perform amplitude
estimation: Grover's algorithm~\cite{Grover97} is to amplitude
amplification what approximate counting is to amplitude estimation.

\begin{theorem}\label{thm:ampleval}
Replacing ${\mathbf G}_F$ in ${\mathbf C}_F$ of algorithm {\bf Count} by
${\mathbf Q} = {\mathbf Q}({\mathcal A},\chi,-1,-1)$
and also modifying Step~6 so that the algorithm outputs
$\tilde a = \sin^2(\tilde{f} \pi/P)$,
$\textup{\textbf{Count}}(F,P)$ with $P \geq 4$
 will output $\tilde{a}$ such that
\[|a - \tilde{a}| < \frac{2 \pi}{P} \sqrt{a} + \frac{\pi^2}{P^2}\]
with probability at least  $8/\pi^2$.
\end{theorem}

In Theorems~\ref{thm:count} and~\ref{thm:ampleval},
parameter~$P$ allows us to balance the desired accuracy of the
estimate with the running time required to achieve~it.
We will now look at different choices for $P$ and
analyse the accuracy of the answer.
To~obtain $t$ up to a few standard deviations,
apply the following corollary of Theorem~\ref{thm:count}.

\begin{corollary}\label{thm:std}
Given a Boolean function $F:\{0,\dots,N-1\} \rightarrow \{0,1\}$
with $t$ as defined above,
{\bf Count}$(F,c \sqrt{N}\,)$ outputs an estimate~$\tilde{t}$ such that
\[|t-\tilde{t}|< \frac{2 \pi}{c} \sqrt{t} + \frac{\pi^2}{c^2}\]
with probability at least
$8/\pi^2$ and requires exactly $c \sqrt{N}$ evaluations of $F$.
\end{corollary}

The above corollary states that some accuracy can be achieved
with probability $8/\pi^2$.
This means that, as usual, the success probability
can be boosted exponentially close to~1 by
repetition.
We will denote by {\bf Maj}$(k,{\bf Count})$ an algorithm
that performs $k$ evaluations of {\bf Count} and outputs the
majority answer.
To~obtain an error probability
smaller than $1/2^n$, one should choose $k$ in $\Omega(n)$.

If one is satisfied in counting up to a constant relative error,
it would be natural to call {\bf Count} with $P = c \sqrt{N/t}$\;\!,
but we need to use the following strategy because $t$ is precisely
what we are looking for.

\medskip\noindent $\textbf{CountRel}(F,c)$
\begin{enumerate}
\item  $P \leftarrow 2$
\item  Repeat
\begin{enumerate}
 \item  $P \leftarrow 2 P$
 \item  $\tilde{f} \leftarrow${\bf  Maj}$({\Omega(\log \log N )},
  ${\bf Count}$ (F,P))$
\end{enumerate}
\item  Until  $\tilde{f} > 1 $
\item  Output {\bf Count}$(F,c P)$
\end{enumerate}
\medskip

Note that in the main loop the algorithm calls {\bf Count}
to obtain $\tilde{f}$ and not $\tilde{t}$.

\begin{corollary}\label{thm:rel}
Given $F$ with $N$ and $t$ as defined above,
{\bf CountRel}$(F,c)$ outputs an estimate~$\tilde{t}$ such that
\[|t-\tilde{t}|< t/c \]
with probability at least~$\frac34$,
using
an expected number of
\mbox{$\Theta( (c +\log \log N)\sqrt{N/t}\,)$}
evaluations of~$F$.
\end{corollary}

\begin{proof}
Suppose for the moment that in Step~\mbox{2(b)}
we always obtain $\tilde{f}$ such that $|f-\tilde{f}| < 1$.
Combining this with Equation~\ref{eqn:f} we see that to obtain
$\tilde{f}>1$, we must have ${P \theta}/ \pi > 1$.
Since $\sin^2 \theta = t/N$, then $P> 2 \sqrt{N/t}$,
so, by~Theorem~\ref{thm:count},
$|t-\tilde{t}|< t \frac{\pi}{c}(1+\frac{\pi}{c})$.
Thus, the core of the main loop will be performed at most
$\log(2 \sqrt{N/t}\,)$ times before $P$ is large enough.
By using $\Omega(\log \log N)$ repetitive calls to {\bf Count} in
Step~\mbox{2(b)},
we know that this will happen with sufficiently high probability,
ensuring an overall success probability of at least~$3/4$.

The expected number of evaluations of~$F$
follows from the fact that
$\sum_{i=1}^{\log( 2 \sqrt{N/t})} (\log \log N) 2^i \in
\Theta \big((\log \log N)\sqrt{N/t}\,\big)$.
\qed
\end{proof}

Of course, to obtain a smaller relative error, the
first estimate can be used in order to call {\bf Count}
with $P$ as large as one wishes. From Theorem~\ref{thm:count},
it is clear that by letting $P$ be large enough,
one can make the absolute error smaller than~1.

\begin{corollary}\label{thm:exact}
Given $F$ with $N$ and $t$ as defined above,
there is an algorithm requiring an expected number of $\Theta(\sqrt{t N}\,)$
evaluations of $F$ that outputs an estimate~$\tilde{t}$
such that $\tilde{t}=t$
with probability at least~$\frac34$
using only space linear in~$\log N$.
\end{corollary}

\begin{proof}
By Theorem~\ref{thm:count}, if $P> \pi(2+\sqrt{6}\,) \sqrt{tN}$, the
error in the output of {\bf Count} is likely to be smaller than~$1/2$.
Again we do not know $t$, but we already know how to estimate it.
By calling first {\bf Count}$(F,\sqrt{N}\,)$ a few times, we obtain an
approximation $\tilde{t}$ such that
$|t-\tilde{t}| < 2 \pi \sqrt{t} + \pi^2$
with good probability.
Now, assuming the first estimate was good, calling {\bf Count}$(F, 20
\sqrt{\tilde{t}N}\,)$ we obtain $\tilde{t'}=t$ with a probability of at least
$8/\pi^2$.
Thus,
obtaining an overall success probability of at least~$3/4$.
\qed
\end{proof}

It follows from a new result of Beals, Buhrman, Cleve, Moska and
de~Wolf~\cite{BBCMW98} that any quantum algorithm capable of deciding
with high probability whether or not a function
\mbox{$F:\{0,\dots,N-1\}\rightarrow \{0,1\}$} is such that
\mbox{$\big| F^{-1}(1) \big| \le t$}, given some \mbox{$0 < t < N/2$},
must query $F$ at least $\Omega(\sqrt{Nt}\,)$ times.
Therefore, our exact counting algorithm is optimal.
Note also that successive applications of Grover's algorithm in which
we strike out the solutions as they are found will also provide
an exact count with high probability,
but at a high cost in terms of additional quantum memory,
that is $\Theta(t)$.

\section*{Acknowledgements}

We are grateful to Joan Boyar, Harry Buhrman,
Christoph D{\"u}rr, Michele Mosca,
Barbara Terhal and Ronald de~Wolf
for helpful comments.
The third author would like to thank M\'elanie Dor\'e Boulet for
her encouragements throughout the realization of this work.

\end{document}